# Mechanism of the electro-optic effect and nonlinear optical susceptibilities of some ferroelectrics: ab initio calculation.

Amirullah M. Mamedov[a], Ekmel Özbay

Nanotechnology Research Center (NANOTAM),
Bilkent University, Bilkent Ankara, 06800, Turkey

**Abstract**

The nonlinear optical properties and electro-optic effects of some oxygen-octahedric ferroelectrics are studied by the density functional theory (DFT) in the local density approximation (LDA) expressions based on first principle calculations without the scissor approximation. We present calculations of the frequency- dependent complex dielectric function $\varepsilon(\omega)$ and the second harmonic generation response coefficient $\chi^{(2)}(-2\omega,\omega,\omega)$ over a large frequency range in tetragonal and rhombohedral phases. The electronic linear electrooptic susceptibility $\chi^{(2)}(-\omega,\omega,0)$ is also evaluated below the band gap. These results are based on a series of the LDA calculation using DFT. Results for $\chi^{(2)}(-\omega,\omega,0)$ are in agreement with the experiment below the band gap and those for $\chi^{(2)}(-2\omega,\omega,\omega)$ are compared with the experimental data where available.

---

[a] Author to whom correspondence should be addressed. Electronic mail: mamedov@bilkent.edu.tr

## 1. INTRODUCTION

Nowadays, nonlinear optics has developed into a field of major study because of rapid advance in photonics [1]. Nonlinear optical techniques have been applied to many diverse disciplines such as condensed matter physics, medicine and chemical dynamics. The development of new advanced nonlinear optical materials for special applications is of crucial importance in technical areas such as optical signal processing and computing, acousto-optic devices and artificial neuro-network implementation. There are intense efforts in experimenting, fabricating and searching for various nonlinear optical materials including ferroelectrics and related compounds. However there is comparatively a much smaller effort to understand the nonlinear optical process in these materials at the microscopic level. The theoretical understanding of the factor that control the figure of merit is extremely important in improving the existing electrooptic (EO) materials and in the search for new ones [2].

There exist a number of calculations for the electronic band structure and optical properties using different methods [3-10]. There is a large variation in the energy gaps, suggesting that the energy band gap depends on the method of the energy spectra calculation. We therefore thought it worthwhile to perform calculations using the density functional theory (DFT) in the local density approximation (LDA) expressions, as implemented within ABINIT package [11] the following convention. Static fields will be labeled by Greek indices ($\alpha,\beta,\ldots$) while we refer to optical fields with Latin symbols ($i,j,\ldots$). To simplify the notation, we will also drop labels such as $\infty$ for quantities that do not involve the response of the ions. Using this convention, we can write $\varepsilon_{ij}$ and $\varepsilon_{\alpha\beta}$, respectively, for the optical and static dielectric tensors, respectively, and $r_{ijk}$ ( $R_{ijkl}$ ) for the linear (quadratic) EO tensor that involves two optical and one static electric fields.

In this paper, we describe detailed calculations of the nonlinear optical properties, includes linear ($r_{ijk}$) and quadratic ($R_{ijkl}$) electro-optic tensors for some $ABO_3$ ferroelectrics. It is well known that the family of the oxygen-octahedral crystals ($ABO_3$, where A=Sr, Ba, Li, K, Pb, Bi, and other elements , B=Ti, Nb, Ta, and other d-transition elements) is one of the most important and numerous groups of ferroelectrics. The structure of these crystals is a combination of oxygen octahedra, in the centers and voids of which other ions are located. The family of $ABO_3$ ferroelectrics has three basic structures:

1. Perovskite (simple and distorted)

2. Trigonal pseudoilmenite,

3. Potassium tungsten bronze

The ferroelectric oxides have a number of properties that make them attractive for use in nonlinear optical devices. Further investigations of these compounds have shown them to be promising for the purposes of the optical reduction of information. As a class of compounds they have wide bandgaps, large electro-optical (EO) and nonlinear optical (NLO) coefficients, high static dielectric constants, and the possibility of sustaining a spontaneous polarization. The interest in these compounds, however, is not restricted to applications only. The presence of the $BO_6$ octahedron with different B-O bonds in $ABO_3$ and the displacement of the B-ion in the octahedron in the course of phase transitions lead to changes in many of the macroscopic and microscopic parameters of these crystals. A study of the role of the $BO_6$ octahedron can cast light on the many physical phenomena that take place in $ABO_3$. All of these have motivated investigations of the linear and nonlinear optical properties of the $ABO_3$ ferroelectrics.

Our aim in this study is to understand the origin of the $\chi^{(2)}(\omega)$, $R_{ijkl}$ and $r_{ijk}$ in these materials as well as to study the trends with moving from Ti to Nb and Nb to Ta (Ba→Li→K). and also to develop the relation between the nonlinear optical properties of $ABO_3$ ferroelectrics and their electronic band structure.

Our paper is organized as follows. In sec.2, we describe the methodology, structure and computational details. In sec.3, we describe the computation of the nonlinear optical susceptibilities and linear and quadratic EO tensors. In sec.4, we illustrate the validity of the formalism by applying methodology and theory (see sec.2 and sec.3) to $ABO_3$ ferroelectrics. Some of the tensor that we consider in this work depends on static electric fields: They include contributions of both the electrons and ions. Other quantities imply only the response of the valence electrons: They are defined for the frequencies of the electric fields high enough to get rid of the ionic contributions but sufficiently low to avoid electronic excitations.

## 2. COMPUTATIONAL DETAILS

The nonlinear optical properties of $ABO_3$ were theoretically studied by means of first principles calculations in the framework of density functional theory (DFT) and based on the local density approximation (LDA) [11] as implemented in the ABINIT code [8,12]. The selfconsistent norm-conserving pseudopotentials are generated using Troullier-Martins scheme [13] which is included in the Perdew-Wang [14] scheme as parameterized by Ceperly and Alder [15]. For calculations, the wave functions were expanded in plane waves up to a kinetic-energy cutoff of 40 Ha ($LiNbO_3$ and $LiTaO_3$), 35 Ha (tetragonal and rhombohedra $KNbO_3$), 38 Ha ($BaTiO_3$). The Brillouin zone was sampled using a 6x6x6 the Monkhorst-Pack [16] mesh of special k points. Rhombohedral position coordinates of $LiNbO_3$ and $LiTaO_3$ using both experimental value [17, 18] were calculated to relate to the hexagonal coordinates given in the literature by the transformation [19]. The coordinates of $KNbO_3$ [20] and $BaTiO_3$ [21] are reported in Table I. All calculations of $ABO_3$ have been used with the experimental lattice constants and atomic positions. The lattice constants and atomic positions are given in Table I. The coordinates of the other atoms can easily be obtained by using the symmetry operations of the space groups. These parameters were necessary to obtain converged results in the nonlinear optical properties.

TABLE I. Atomic positions and lattice constants of some $ABO_3$ crystals.

| Phase | Space Group | Lattice Parameters (Å) | Atom | Position |
|---|---|---|---|---|
| $LiNbO_3$ Ferroelectric (Rhombohedral) | R3c | a = b = c = 5.4944 | Li<br>Nb<br>O | (0.2829, 0.2829, 0.2829)<br>(0.0, 0.0, 0.0)<br>(0.1139, 0.3601, -0.2799) |
| $LiTaO_3$ Ferroelectric (Rhombohedral) | R3c | a = b = c = 5.4740 | Li<br>Ta<br>O | (0.2790, 0.2790, 0.2790)<br>(0.0, 0.0, 0.0)<br>(0.1188, 0.3622, -0.2749) |
| $KNbO_3$ Ferroelectric (Tetragonal) | P4mm | a = b = 3.9970<br>c = 4.0630 | K<br>Nb<br>O(1)<br>O(2) | (0.0, 0.0, 0.023)<br>(0.5, 0.5, 0.5)<br>(0.5, 0.5, 0.04)<br>(0.5, 0.0, 0.542) |
| $KNbO_3$ Ferroelectric (Rhombohedral) | R3m | a = b = c = 4.0160 | K<br>Nb<br>O(1)<br>O(2) | (0.0112, 0.0112, 0.0112)<br>(0.5, 0.5, 0.5)<br>(0.5295, 0.5295, 0.0308)<br>(0.5295, 0.0308, 0.5295) |
| $BaTiO_3$ Ferroelectric (Tetragonal) | P4mm | a = b = 3.9909<br>c = 4.0352 | Ba<br>Ti<br>O(1)<br>O(2) | (0.0, 0.0, 0.0)<br>(0.5, 0.5, 0.5224)<br>(0.5, 0.5, -0.0244)<br>(0.5, 0.0, 0.4895) |

## 3. LINEAR AND NONLINEAR OPTICAL RESPONSE

*3.1 Linear Optical Response*

It is well known that the effect of the electric field vector, $E(\omega)$, of the incoming light is to polarize the material. In an insulator the polarization can be expressed as a Taylor expansion of the $E(\omega)$

$$P^i(\omega) = P_s^i + \sum_{j=1}^{3} \chi_{ij}^{(1)}(-\omega,\omega)E^j(\omega) + \sum_{j,l=1}^{3} \chi_{ijl}^{(2)} E^j(\omega)E^l(\omega) + ... \quad (1)$$

where $P_s^i$ is the zero field (spontaneous) polarization, $\chi_{ij}^{(1)}$ is the linear optical susceptibility tensor and is given by ref. [22].

$$\chi_{ij}^{(1)}(-\omega,\omega) = \frac{e^2}{\hbar\omega} \sum_{n,m,\vec{K}} f_{nm}(\vec{K}) \frac{r_{nm}^i(\vec{K}) r_{mn}^j(\vec{K})}{\omega_{mn}(\vec{K}) - \omega} = \frac{\varepsilon_{ij}(\omega) - \delta_{ij}}{4\pi} \quad (2)$$

where n, m denote energy bands, $f_{mn}(\vec{K}) = f_m(\vec{K}) - f_n(\vec{K})$ is the Fermi occupation factor, and $\Omega$ is the normalization volume.

$\omega_{mn}(\vec{K}) \equiv [\omega_m(\vec{K}) - \omega_n(\vec{K})]$ is the frequency difference and $\hbar\omega_n(\vec{K})$ is the energy of band n at wave vector $\vec{K}$. The rijk are the matrix elements of the position operator and are given by

$$r_{nm}^i(\vec{K}) = \frac{v_{nm}^i(\vec{K})}{i\omega_{nm}}; \omega_n \neq \omega_m$$

$$r_{nm}^i(\vec{K}) = 0, \omega_n = \omega_m \quad (3)$$

where $v_{nm}^i(\vec{K}) = [P_{nm}^i(\vec{K})/m]$, m is the free electron mass, and $P_{nm}$ is the momentum matrix element. $\chi_{ijl}^{(2)}$ the second-order nonlinear susceptibility tensor and is discussed in sec.4. As can be seen from equation (2), the dielectric function $\varepsilon_{ij}(\omega) = [1 + 4\pi\chi_{ij}^{(1)}(-\omega,\omega)]$ and the imaginary part of $\varepsilon_{ij}(\omega)$, $\varepsilon_2^{ij}(\omega)$ is given by

$$\varepsilon_2^{ij}(\omega) = \frac{e^2}{\hbar\pi} \sum_{nm} \int d\vec{K} f_{nm}(\vec{K}) \frac{v_{nm}^i(\vec{K}) v_{mn}^j(K)}{\omega_{mn}^2} \delta(\omega - \omega_{mn}(K)) \quad (4)$$

The real part of $\varepsilon_{ij}(\omega), \varepsilon_1^{ij}(\omega)$, can be obtained by using Kramers-Kronig transformation

$$\varepsilon_1^{ij}(\omega) - 1 = \frac{2}{\pi} P \int_0^\infty \frac{\omega' \varepsilon_2^{ij}(\omega')}{\omega'^2 - \omega^2} d\omega' \quad (5)$$

As the Kohn-Sham equations only determine the ground-state properties, hence the unoccupied conduction bands have no physical significance. If they are used as single-particle states in the optical calculation of semiconductors, a band gap problem comes into existence: The absorption starts at an excessively low low energy [18]. In order to remove the deficiency the many-body effects must be included in calculations of response functions. In order to take into account the self-energy effects, the scissors approximation is generally used [23]. In the calculation of the optical response in present work we have used the standard expression for $\varepsilon_{ij}(\omega)$ (see equations (4) and (5)).

*3.2 Nonlinear Response*

The general expression of the nonlinear optical susceptibility depends on the frequencies of the $E(\omega)$. Therefore, in present context of the (2n+1) theorem applied within the LDA to DFT we get an expression for the second order susceptibility [22-26]. As the sum of the three physically different contributions

$$\chi^{(2)}_{ijl}(-\omega_\beta,-\omega_\gamma,\omega_\beta,\omega_j) = \chi^{"}_{ijl}(-\omega_\beta,-\omega_\gamma;\omega_\beta,\omega_j) + \eta^{"}_{ijl}(-\omega_\beta,-\omega_j,\omega_\beta,\omega_j) + i\frac{\sigma^{"}_{ijl}(-\omega_\beta,-\omega_j,\omega_\beta,\omega_\gamma)}{\omega_\beta+\omega_\gamma} \quad (6)$$

That includes contributions of interband and intraband transitions to the second order susceptibility. The first term in equation (6) describes contribution of inter band - transitions to second order susceptibility. The second term represents the contribution of intraband transitions to second order susceptibility and the third term is the modulation of interband terms by intrabands terms. We used this expression to calculate the nonlinear response functions of $ABO_3$ ferroelectrics.

*3.3 Principal Refractive Indices Calculation*

The principal refractive indices, $n_i$, can be computed as a square root of the eigenvalues of the optical dielectric tensor. At finite temperature, T, we can write $<\varepsilon_{ij}(u_r,\eta)> = \delta_{ij} + 4\pi<\mu^1_{ij}(u_r,\eta)>$ where <...> refers to the average value at a given T. Let us write $u_r$ and $\eta$ as $u_r = <u> + \delta u_r$ and $\eta = <\eta> + \delta\eta$, where $\delta u_r$ and $\delta\eta$ denote the deviations from average values (here, $u_r$ - the ionic degree of freedom in r unit cell, $\eta$ - the macroscopic strains). If we develop $<\chi^{(1)}_{ij}(u_r,\eta)>$ as a Taylor expansion about the paraelectric structure, we can separate the terms depending on $<u>$ and $<\eta>$ only from those involving also $\delta u_r$ and $\delta\eta$. At a finite temperature, the dielectric susceptibility can therefore be expressed as

$$<\chi^{(1)}_{ij}(u_r,\eta)> = \chi^{(1)}_{ij}(<u>,<\eta>) + <\chi^{(1)}_{ij}(<u>,<\eta>,\delta u_r,\delta\eta)> \quad (7)$$

The first term describes the variations of $\chi_{ij}^{(1)}$ due to the averaged crystal lattice distortions. It is responsible for the discontinuity of $n_i$ at the phase transition in ferroelectrics such as BaTiO$_3$. The second term represents the variations of $\chi_{ij}^{(1)}$ due to thermal fluctuations and to their correlations [22]. It determines the variations of $n_i$ in the paraelectric phase. This term is difficult to compute in practice. However, in the usual ferroelectric such as BaTiO$_3$ or KNbO$_3$, the variations of $n_i$ in the paraelectric phase are small compared to their variation at the phase transition. Following ref. [22] we neglect the second term in equation (7) since we are interested in the variation of $n_i$ below the phase transition temperature (Tc) where we expect the first term to dominate. The linear EO effect is related to the first order change of the optical dielectric tensor induced by a static or low frequency electric field (E).

*3.4 Electro-Optic Tensor*

The optical properties of material usually depend on external parameters such as the temperature, electric or magnetic fields or mechanical constraints (stress, strain). Now we consider the variations of the refractive index induced by a static or low-frequency electric field E. Band theoretical expression for the EO effect have been given in [22]. The first principles calculations of this type not only neglect important excitonic contributions to $\varepsilon_2^{ij}(\omega)$ in ionic crystals, but they also are not presently feasible even for such relatively simple compounds as TiO$_2$. For this reason, theories of the EO effect require some approximations and parameterizations within the framework of either general quantum theories or physically appealing simplified models, like quantum anharmonic oscillator model of Robinson [27], where the octapole moment of the ground state (valance band) charge density serves as a measure of acentricity. When the reasonably accurate wavefunctions are available, this theory provides a formalism for the computation of the EO effect. It should be noted that the ground state theory of Robinson does not display explicitly the importance of the interband transitions to electrooptics. Although, as emphasized by Robinson and other investigators [27], knowledge of the excited states (conduction bands) is contained within the exact ground state (valance bands) wavefunctions, there appears to be no straightforward procedure for establish in the accuracy of the conduction bands generated by the necessarily approximate ground state. A connection between the energy band approach and the ground state (moment) theory can be obtained by an equation

$$T_{K,ji} \approx -(\frac{a_0 e}{24\pi^2 P_K^e}) P \int_{\omega_g^1}^{\infty} \frac{\omega' \Delta\varepsilon_{2,ij}(\omega')}{(\omega'^2 - \omega^2)} \qquad (8)$$

Thus the octapole moment T$_{k,ji}$ per valence electron is related to a Kramers-Kronig integral over the polarization-induce changes in the fundamental $\varepsilon_2^{ij}(\omega)$ spectrum. ($P_K^e$ is the field induced electronic polarization) To summarize, the fundamental theories of the EO effect involving either k-space integrations are presently capable of quantitatively predicting the magnitude of EO coefficients, because the first principles calculations using DFT in the LDA expressions are successful. Below we review this approach to linear and quadratic EO effects.

### 3.4.1 Linear Electro-Optic Effect

At linear order, these variations are described by the linear electro-optical (EO) coefficients (Pockels effect).

$$\Delta(\varepsilon^{-1})_{ij} = \sum_{k=1}^{3} r_{ijk} E_k \tag{9a}$$

where $(\varepsilon^{-1})_{ij}$ is the inverse of the electronic dielectric tensor and $r_{ijk}$ the EO tensor. Within the Born-Oppenheimer approximation, the EO tensor can be expressed as the sum of the three contributions: a bare electronic part $r_{ijk}^{el}$ an ionic contribution $r_{ijk}^{ion}$ and a piezoelectric contribution $r_{ijk}^{piezo}$. The electronic part is due to an interaction of $E_k$ with the valence electrons when considering the ions artificially as clamped at their equilibrium positions. It can be computed from the nonlinear optical coefficients. As can be seen from equation (6) $\chi_{ijl}^{(2)}$ defines the second order change of the induced polarization with respect to $E_k$. Taking the derivation of equation (9) we also see that $\chi_{ijl}^{(2)}$ defines the first-order change of the linear dielectric susceptibility, which is equal to $(1/4\pi)\Delta\varepsilon_{ij}$. Since the EO tensor depends on $\Delta(\varepsilon^{-1})_{ij}$ rather than $\Delta\varepsilon_{ij}$, we have to transform $\Delta\varepsilon_{ij}$ to $\Delta(\varepsilon^{-1})_{ij}$ by the inverse of the zero-field electronic dielectric tensor [8].

$$\Delta(\varepsilon^{-1})_{ij} = -\sum_{m,n=1}^{3} \varepsilon_{im}^{-1} \Delta\varepsilon_{mn} \varepsilon_{nj}^{-1}$$

(9b)

Using equation (9b) we obtain the following expression for the electronic EO tensor:

$$r_{ijk}^{el} = -8\pi \sum_{l,l'=1}^{3} (\varepsilon^{-1})_{il} \chi_{ll'k}^{(2)} (\varepsilon^{-1})_{l'j} \tag{10}$$

Equation (10) takes a simpler from when expressed in the principal axes of the crystal under investigation [8]:

$$r_{ijk}^{el} = -\frac{8\pi}{n_i^2 n_j^2} \chi_{ijk}^{(2)} \tag{11}$$

where $n_i$ coefficients are the principal refractive indices.

The origin of ionic contribution to the EO tensor is the relaxation of the atomic positions due to the applied electric field $E_k$ and the variations of the $\varepsilon_{ij}$ induced by these displacements. It can be computed from the Born effective charge $Z^*_{k,\alpha,\beta}$ and the $\partial \chi_{ij} / \partial T_{k\alpha}$ coefficients introduced in [8].

The ionic EO tensor can be computed as a sum over the transverse optic phonon modes at $\vec{q} = 0$.

$$r_{ijk}^{ion} = -\frac{4\pi}{\sqrt{\Omega} n_i^2 n_j^2} \sum_m \frac{\alpha_{ij}^m P_{mk}}{\omega_m^2} \quad (12)$$

where $\alpha^m$ is the Raman susceptibility of mode m and $P_{m,k}$ the mode polarity

$$P_{mk} = \sum_{k',\beta} Z^*_{k',k\beta} u_m(k'\beta) \quad (13)$$

which is directly linked to the make oscillator strength

$$S_{m,\alpha\beta} = P_{m,\alpha} P_{m\beta} \quad (14)$$

For simplicity, we have expressed equation (14) in the principal axes while a more general expression can be derived from equation (10).

Finally, the piezoelectric contribution is due to the relaxation of the unit cell shape due to the converse piezoelectric effect [8]. It can be computed from the elasto-optic coefficients $P_{ij\mu\nu}$ and the piezoelectric strain coefficients $d_{k\mu\nu}$:

$$r_{ijk}^{piezo} = \sum_{\mu,\nu=1}^{3} P_{ij\mu\nu} d_{k\mu\nu} \quad (15)$$

In the discussion of the EO effect, we have to specify whether we are dealing with strain-free (clamped) or stress-free (unclamped) mechanical boundary conditions. The clamped EO tensor $r_{ijk}^{\eta}$ takes into account the electronic and ionic contributions but neglects any modification of the unit cell shape due to the converse piezoelectric effect [8].

$$r_{ijk}^{\eta} = r_{ijk}^{el} + r_{ijk}^{ion} \quad (16)$$

Experimentally, it can be measured for the frequencies of $E_k$ high enough to eliminate the relaxations of the crystal lattice but low enough to avoid excitations of optical phonon modes (usually above ~ 102 MHz). To compute the unclamped EO tensor $r_{ijk}^{\sigma}$ we added the piezoelectric contribution to $r_{ijk}^{\eta}$. In the noncenterosymmetric phases of $ABO_3$ the EO tensor has four independent elements r13, r33, r22, r15 = r42. In contrast to the dielectric tensor, the EO coefficients can either be positive or negative. The sign of these coefficients is often difficult to measure experimentally. Moreover, it depends on the choice of the Cartesian axes. The z-axis is along the direction of the spontaneous polarization and the y–axis lies in a mirror plane. The z and y–axes are both piezoelectric.

Their positive ends are chosen in the direction that becomes negative under compression. The orientation of these axes can easily be found from pure geometrical arguments. Our results are reported in the Cartesian axes where the piezoelectric coefficients $d_{22}$ and $d_{33}$ are positive. These coefficients, as well as their total and electronic part, are reported in Table 2. All EO coefficients are positive as is the case for the noncentro-symmetric phases [8], the phonon modes that have the strongest overlap with the soft mode of the paraelectric phase dominate the amplitude to the EO coefficients. Moreover, the electronic contributions are found to be quite small. All of our investigation of EO coefficients of $ABO_3$ shows a good agreement and also between our results and earlier experimental investigations.

TABLE II. EO tensors of some $ABO_3$ crystals

| Crystals | Symmetry Class | Linear | EO coefficients x $10^{-7}$(esu) | | | Quadratic EO coefficients, x $10^{-12}$ (esu) | Total |
|---|---|---|---|---|---|---|---|
| | | | Electronic | Total | Exp. | | |
| $BaTiO_3$ | 4mm | $r_{13}$ | 0.358 | 1.653 | 3.06 [28] | $R_{11}$ | 8.2 |
| | | $r_{33}$ | 0.505 | 3.570 | 12.18 [28] | $R_{12}$ | 1.7 |
| | | $r_{51} = r_{42}$ | 0.399 | 19.533 | | $R_{33}$ | 12.5 |
| $KNbO_3$ | 4mm | $r_{13}$ | 0.288 | 1.279 | | $R_{11}$ | 91.3 |
| | | $r_{33}$ | 1.029 | 5.117 | | $R_{12}$ | 20.7 |
| | | $r_{51} = r_{42}$ | 0.288 | 1.279 | | $R_{44}$ | 12.5 |
| | 3m | $r_{13}$ | 0.569 | 3.417 | | | |
| | | $r_{33}$ | 0.942 | 6.276 | | | |
| | | $r_{51} = r_{42}$ | 0.623 | 3.459 | | | |
| | | $r_{22}$ | 0.254 | 1.333 | | | |
| $LiNbO_3$ | 3c | $r_{13}$ | 0.230 | 1.756 | 2.58 [28] | | |
| | | $r_{33}$ | 0.082 | 6.085 | 9.24 [28] | | |
| | | $r_{51} = r_{42}$ | 0.236 | 1.879 | 8.40 [28] | | |
| | | $r_{22}$ | 0.002 | 0.402 | 1.02 [28] | | |
| $LiTaO_3$ | 3c | $r_{13}$ | 0.092 | 3.513 | 2.52 [28] | | |
| | | $r_{33}$ | 0.718 | 5.151 | -0.06 [28] | | |
| | | $r_{51} = r_{42}$ | 0.091 | 1.105 | 9.15 [28] | | |
| | | $r_{22}$ | 0.039 | 0.132 | 6.00 [28] | | |

TABLE III. Total, Intraband and Interband values of Re $\chi_{ijk}^{(2)}$ (0), ($10^{-9}$ esu)

| Compound | Re $\chi_{ijk}^{(2)}$ (0) | $\chi_{113}^{(2)} = \chi_{131}^{(2)}$ | $\chi_{222}^{(2)}$ | $\chi_{311}^{(2)}$ | $\chi_{333}^{(2)}$ |
|---|---|---|---|---|---|
| BaTiO$_3$ | Inter($\omega$) | -0.2319 | - | 0.0217 | 0 8522 |
| | Inter($2\omega$) | -0.5912 | - | -0.4714 | -2.3513 |
| | Intra($\omega$) | 0.0272 | - | -0.2121 | -0.3977 |
| | Intra($2\omega$) | 1.0027 | - | -0.0978 | -0.2230 |
| | Total | 0.2068 | - | -0.7581 | -2.1197 |
| KNbO$_3$ | Inter($\omega$) | 0.0339 | - | -0.9580 | 0.3651 |
| | Inter($2\omega$) | 0.6190 | - | 1.5967 | -1.2382 |
| | Intra($\omega$) | 0.0858 | - | 0.2758 | 0.0 53 |
| | Intra($2\omega$) | -1.2742 | - | 0.8832 | 2.0256 |
| | Total | -0.5355 | - | 1.7997 | 1.2077 |
| LiNbO$_3$ | Inter($\omega$) | -0.0629 | -0.1208 | -0.1655 | 0.5775 |
| | Inter($2\omega$) | -1.7209 | 0.7369 | -1.7243 | -1.750 |
| | Intra($\omega$) | -0.2768 | -0.0170 | -0.1463 | -0.0783 |
| | Intra($2\omega$) | 2.8817 | -0.6579 | 3.0059 | 2.0675 |
| | Total | 0. 214 | -0.0588 | 0.9718 | 0.3962 |
| LiTaO$_3$ | Inter($\omega$) | -0.0932 | -0.1182 | -0.2068 | 0.1451 |
| | Inter($2\omega$) | -0.0731 | 0.3972 | -0.0425 | -0.6212 |
| | Intra($\omega$) | -0.1124 | -0.0376 | 0.0165 | 0.0173 |
| | Intra($2\omega$) | 0.5419 | -0.3019 | 0.6362 | 0.6414 |
| | Total | 0.2634 | -0.0606 | 0.4030 | 0.1825 |

### 3.4.2. Quadratic Electro-Optic Effect (Kerr Effect)

For our knowledge of the energy band structure and polarization induced energy band changes, we can compute the quadratic EO (Rijkl) coefficients. This model applies in the zero-strain limit, and as consequence, we compute the "clamped" coefficients, defined by [27].

$$\Delta(\frac{1}{n^2})_{ij} = \sum_{K,l=1}^{3} R_{ijkl} E_K E_e \quad (17)$$

for the $ABO_3$ ferroelectrics in their centrosymmetric phase. The refractive index change $\Delta n$ resulting from polarization induced band changes ($(e/hc)\Delta E_g$) can be related to the EO $R_{ijkl}$ coefficients and the polarization-potential tensor concept introduced in [27], as

$$\Delta n = \frac{1}{2} n^3 R P^2 \quad (18)$$

(for the different geometry and symmetry of the compounds R → $R_{11}$, $R_{12}$, $R_{44}$, and so on). The relationship between $r_{ijk}$ of a polarized crystal and $R_{ijkl}$ of an unpolarized crystal were derived in [27] for all crystals symmetries from $O_h$ to $C_{4v}$ (for four-fold octahedral axes) and $C_{3v}$ (for three-fold octahedral axes).

## 4. RESULTS AND DISCUSSION

The calculation of nonlinear optical properties is much more complicated than the same procedure in the linear case. The difficulties concern both the numerical and the analytical solutions. The k-space integration in expression (6) has to be performed more carefully using a generalization of methods [24-26]. More conduction bands have to be taken into account to reach the same accuracy. The fact that the SHG coefficients are related to the optical transitions has remarkable consequences. First of all, we note that the equations for SHG consist of a number of resonant terms. In this sense the imaginary part, $Im\chi^{(2)}(-2\omega,\omega,\omega)$ resembles the $\epsilon_2(\omega)$ and provides a link to the band structure. The difference, however, is that whereas in $\epsilon_2(\omega)$ only the absolute value of the matrix elements squared enters, the matrix elements entering the various terms in $\chi^{(2)}$ are more varied.

They are in general complex and can have any sign. Thus, $Im\chi^{(2)}(-2\omega,\omega,\omega)$ can be both positive and negative. Secondly, there appear both resonances when $2\omega$ equals an interband energy and when $\omega$ equals an interband energy. Figures (1-5) shows the $2\omega$ and single $\omega$ resonances contributions to $Im\chi^{(2)}(-2\omega,\omega,\omega)$ compared to $\epsilon_2(\omega)$ (figure 6) for a number of $ABO_3$. They clearly show a greater variation from high symmetry to the lowest symmetry than the linear optic function. In some sense they resemble a modulated spectrum. Third, we note that the $2\omega$ resonances occur at half the frequency corresponding to the interband transition. Thus, the incoming light need not be as high in the UV to detect this higher lying interband transition. This is important for wide band gap materials like $ABO_3$ compounds where laser light sources reaching the higher interband transitions are not available. Nevertheless, one still needs to be able to detect the corresponding $2\omega$ signal in the UV. Unfortunately the intrinsic richness of $\chi^{(2)}$ spectra remains largely to be explored experimentally and we are not aware of any attempts to measure both the real and imaginary parts of the these spectral functions as one standard does in linear optics. We also calculated the real part (total, intra and inter components) of the SHG susceptibilities Re $\chi^{(2)}_{ijk}$ (0) (Table III). As can be seen from the Table III, the value of $\chi_{311}(0)$ is the dominant component for all $ABO_3$.

It is well known that nonlinear optical properties are more sensitive to small changes in the band structure than the linear optical properties. That is attributed to the fact that the second harmonic response $\chi^{(2)}_{ijk}(\omega)$ contains $2\omega$ resonance along with the usual $\omega$ resonance. Both the $\omega$ and resonances can be further separated into interband and intraband contributions. The structure in $\chi^{(2)}_{ijk}(\omega)$ can be understood from the structures in $\epsilon_2(\omega)$. Our calculations for $\epsilon_2(\omega)$ give two fundamental oscillator bands at ~6 and ~10 eV which correspond to the optical transitions from the valance bands to the conduction band, formed by the d orbits of the B (Ti,Nb,Ta) atoms and consisting of two subbands. It is well known that the $\epsilon_2(\omega)$ function computed from moments ($\vec{p}$) appear to be very sensitive to the ab initio parameters and seem to be particularly appropriate to test the electronic band structure. In $ABO_3$ perovskites the two peak present in the experimental reflectivity data are obtained in theoretical curves only when the interband transition moments varied with respect to the energies and $\vec{k}$ wave vectors. In this computation on $ABO_3$, compounds many parameters have been borrowed from existing computations have been neglected, explaining some discrepancies between theory and experiments [9-10, 29-35]. The structure 2-6 eV in $\chi^{(2)}_{ijk}(\omega)$ is associated with interference between a $\omega$ and $2\omega$ resonances, while the structure above 6 eV is due to mainly to $\omega$ resonance. In Figure 1-5 we show the $2\omega$ interband and intraband contributions for $ABO_3$ compounds. Also given is their decomposition into intra- and interband contributions. They are arranged so as to move the Ba → K → Li, Ti → Nb → Ta trends obvious. For example $\chi^{(2)}$ obviously increases when going from Ba to K and Li and from Ti to Nb. Unfortunately, the agreement between theory and experiment is by no means perfect [36].

Note that the interband part are negative in all cases and in most cases largely compensate the intraband part. The exceptions are the $LiBO_3$ (B=Nb,Ta) compounds in both cases of which interband part is much smaller in magnitude than the intraband part. This quite interesting because unexpected. It raises the question what features in the band structure of these two compounds distinguish them from the other compounds [37,38]. We investigated the reasons for the cancellation of intra- and interband parts by inspecting the corresponding frequency dependent imaginary parts of the $\chi^{(2)}$ ($-2\omega,\omega,\omega$).

First of all, one now sees that the opposite sign of intra- and inter-band parts not only occurs in the static value but also occurs almost energy by energy. This is true over the entire energy range in $BaTiO_3$ and over most of the range (E >1 eV) for other $ABO_3$. The sign of the inter and intraband part are difficult to understand a-priori because a variety of matrix element products comes into play and both $\omega$ and $2\omega$ resonances occur in both the pure interband, and the interband contribution modified by intraband motion when these are further worked out into separate resonance terms. The spectra $\epsilon_2(\omega)$ (figure 6) for the $ABO_3$ compounds are rather similar. They look like the superposition of the spectra of more or less four pronounced oscillators with resonance frequencies close to the M and Z line structures appearing in the $2\omega$ and $\omega$ – terms of the imaginary parts.

As an example of such a prediction the SHG coefficients of $ABO_3$ compounds are given in Table IV. For incident light with a frequency that is small compared to the energy gap. The independent tensor components are listed for $\omega=0$. The comparison with recent experimental values and theoretical calculations [39] are also rather successful where available for the static SHG coefficients of the $ABO_3$ compounds.

TABLE IV. Second order nonlinear optical susceptibilities for some $ABO_3$, ($10^{-7}$ esu).

| Crystals | Symmetry Class | | $d_{15}$ | $d_{22}$ | $d_{31}$ | $d_{33}$ | Ref. |
|---|---|---|---|---|---|---|---|
| $BaTiO_3$ | 4mm | (cal.) | 2.547 | - | 2.547 | 2.885 | |
| | | (exp) | 5.1 | - | 4.71 | 2.04.) | [40] |
| $KNbO_3$ | 4mm | (cal.) | 2.190 | - | 2.190 | 5.322 | |
| | | (cal.) | - | - | -0.299 | -0.818 | [39] |
| | 3m | (cal.) | - | 1.546 | 3.465 | 4.788 | |
| | | (cal.) | - | 0.342 | 0.121 | 0.342 | [39] |
| $LiNbO_3$ | 3c | (cal.) | - | 0.013 | 1.541 | 6.877 | |
| | | (exp.) | - | 0.774 | -1.464 | -10.2 | [40] |
| $LiTaO_3$ | 3c | (cal.) | - | 0.221 | 0.513 | 4.114 | |
| | | (exp.) | - | 0.51 | -0.321 | -4.92.) | [40] |

## 5. CONCLUSION

The linear and nonlinear optical properties for important group of oxygen-octahedron ferroelectrics $ABO_3$ ($LiNbO_3$, $LiTaO_3$, $KNbO_3$ and $BaTiO_3$) have been calculated over a wide energy range. We studied some possible combination of A and B. This allowed us to study the trends in the second order optical response with chemical composition. The results for the zero-frequency limit of second harmonic generation are in agreement with available experimental results. The calculated linear and quadratic electrooptical coefficients for $LiNbO_3$, $LiTaO_3$, $KNbO_3$ and $BaTiO_3$ are also show agreement with recent experimental data in the energy region below the band gap. For all the considered compounds the SHG coefficient $\chi^{(2)}$ is of the order of ~10-7 esu. Our calculations of the SHG susceptibility shows that the intra-band and interband contributions are significantly changes with change B and A – ions.

The authors are grateful to the ABINIT group for the ABINIT project that we used in our computations.

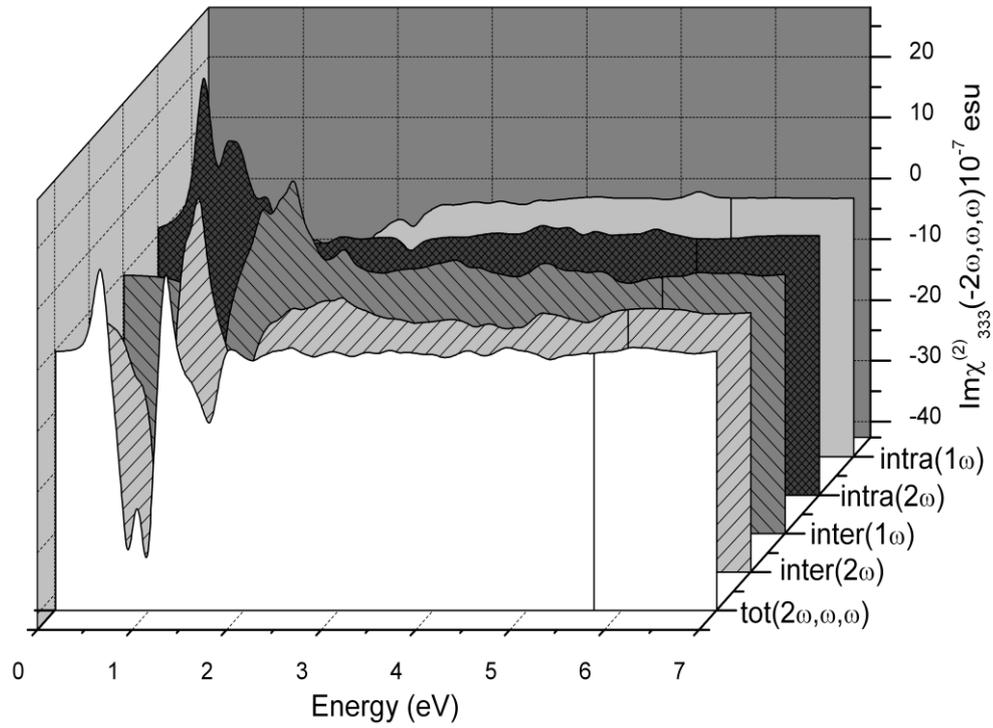

**FIGURE 1**. Second-order susceptibility Im $\chi^{2}_{333}$ (-2ω,ω,ω) for BaTiO$_3$

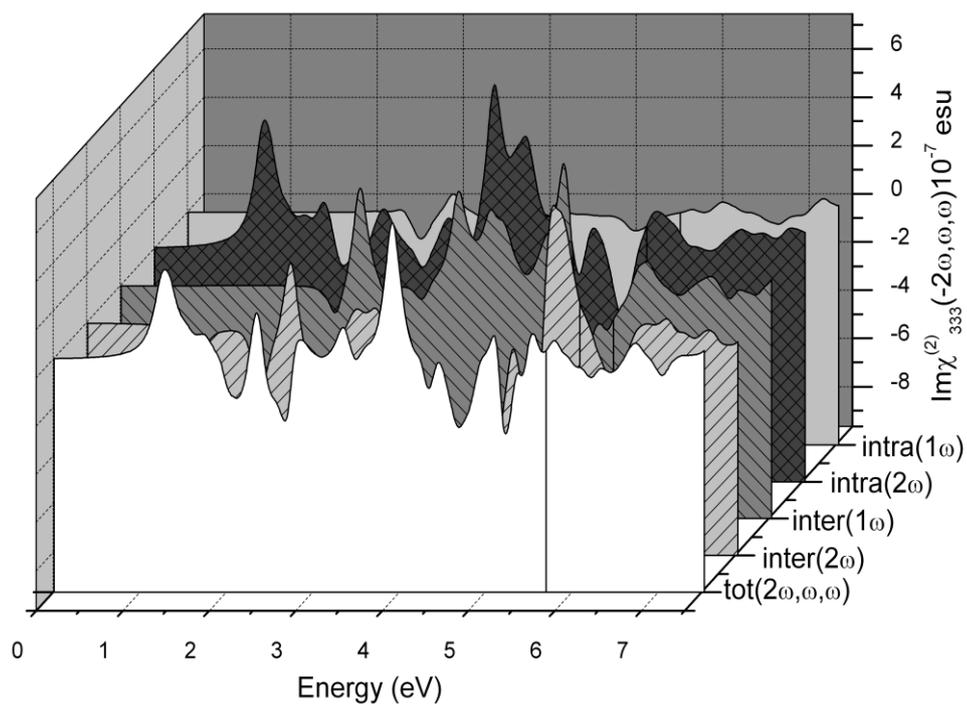

**FIGURE 2**. Second-order susceptibility Im $\chi^2_{333}$ (-2ω,ω,ω) for tetragonal KNbO$_3$

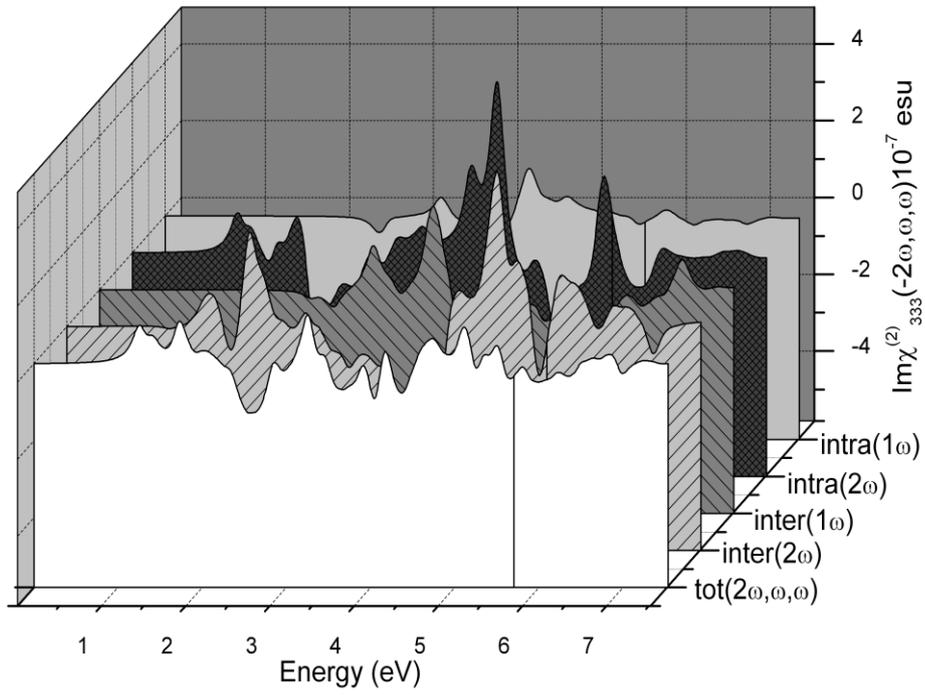

**FIGURE 3**. Second-order susceptibility Im $\chi^{2}_{333}$(-2ω,ω,ω) for rhombohedral $KNbO_3$

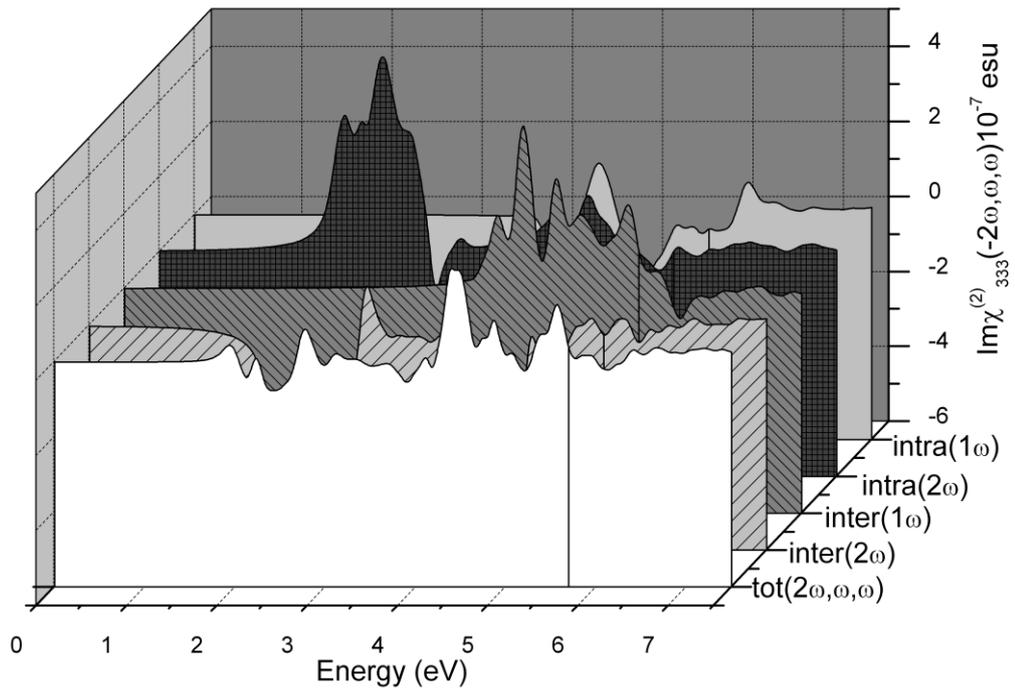

**FIGURE 4.** Second-order susceptibility Im $\chi^{2}_{333}$ (-2ω,ω,ω) for LiNbO$_3$

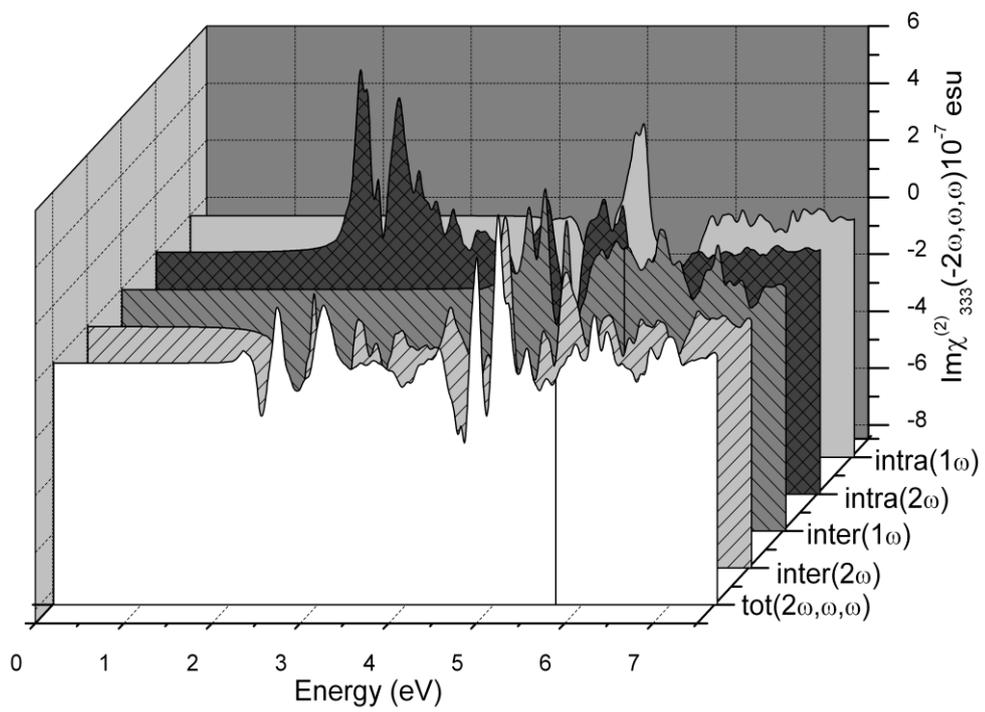

**FIGURE 5**. Second-order susceptibility Im $\chi^{2}_{333}$ (-2ω,ω,ω) for LiTaO$_3$

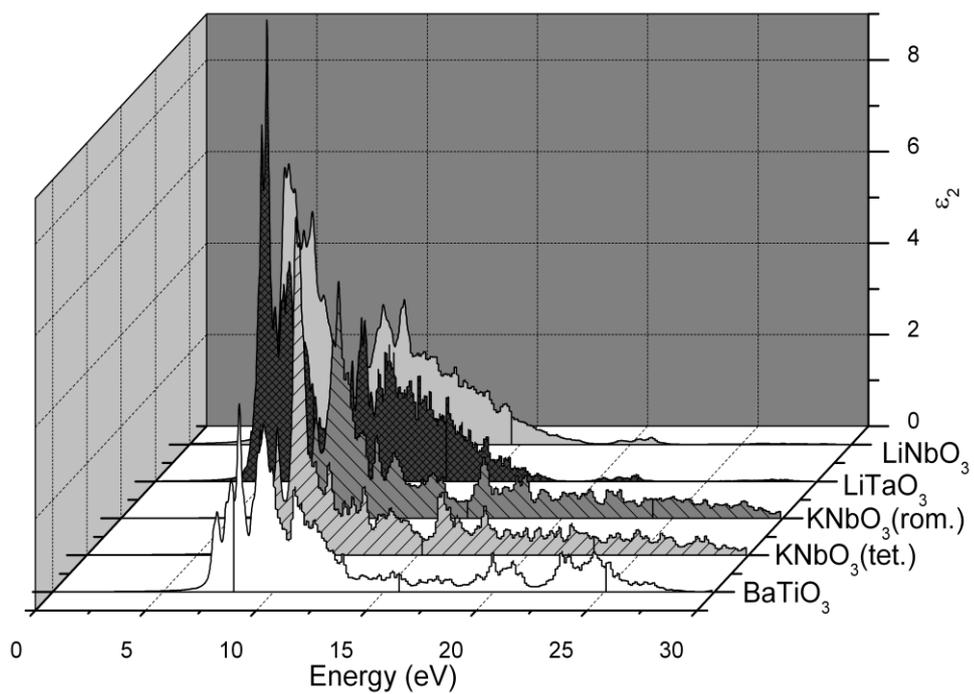

**FIGURE 6**. The calculated imaginary part of z –components of the dielectric function of some ABO$_3$ crystals